\documentclass[prd,floats,floatfix,nofootinbib]{revtex4}
\usepackage{revsymb}
\usepackage{graphicx}
\usepackage{bm}
\usepackage{amsmath}
\usepackage{amssymb}
\usepackage{graphicx}

\begin{document}

\title{ NEWTONIAN APPROACH TO THE MATTER POWER SPECTRUM OF THE
GENERALIZED CHAPLYGIN GAS}

\author{J.C. Fabris\footnote{E-mail: fabris@cce.ufes.br. Present address:
Institut d'Astrophysique de Paris - IAP, Paris, France.}}
\affiliation{Universidade Federal do Esp\'{\i}rito Santo,
Departamento
de F\'{\i}sica\\
Av. Fernando Ferrari, 514, Campus de Goiabeiras, CEP 29075-910,
Vit\'oria, Esp\'{\i}rito Santo, Brazil}

\author{S.V.B.
Gon\c{c}alves\footnote{E-mail: sergio.vitorino@pq.cnpq.br. Present address: Laboratoire d'Annecy-le-Vieux de Physique Theorique - LAPTH,
Annecy-le-Vieux, France.}}
\affiliation{Universidade Federal do Esp\'{\i}rito Santo,
Departamento
de F\'{\i}sica\\
Av. Fernando Ferrari, 514, Campus de Goiabeiras, CEP 29075-910,
Vit\'oria, Esp\'{\i}rito Santo, Brazil}

\author{H.E.S.
Velten\footnote{E-mail: velten@cce.ufes.br}}
\affiliation{Universidade Federal do Esp\'{\i}rito Santo,
Departamento
de F\'{\i}sica\\
Av. Fernando Ferrari, 514, Campus de Goiabeiras, CEP 29075-910,
Vit\'oria, Esp\'{\i}rito Santo, Brazil}

\author{W. Zimdahl\footnote{E-mail: zimdahl@thp.uni-koeln.de}}
\affiliation{Universidade Federal do Esp\'{\i}rito Santo,
Departamento
de F\'{\i}sica\\
Av. Fernando Ferrari, 514, Campus de Goiabeiras, CEP 29075-910,
Vit\'oria, Esp\'{\i}rito Santo, Brazil}

\begin{abstract}
We model the cosmic medium as the mixture of a generalized
Chaplygin gas and a pressureless matter component. Within a
neo-Newtonian approach we compute the matter power spectrum. The
2dFGRS data are used to discriminate between unified models of the
dark sector and different models, for which there is separate dark
matter, in addition to that accounted for by the generalized
Chaplygin gas. Leaving the corresponding density parameters free,
we find that the unified models are strongly disfavored. On the
other hand, using unified model priors, the observational data are
also well described, in particular for small and large values of
the generalized Chaplygin gas parameter $\alpha$.
\end{abstract}

\maketitle

Among the host of models that have been proposed for dark matter
and dark energy over the last years, there are unified models of
the dark sector according to which there is just one dark
component that simultaneously plays the role of dark matter and
dark energy. The most popular proposal along this line is the
Chaplygin gas, an exotic fluid with negative pressure that scales
as the inverse of the energy density \cite{moschella}. This
phenomenologically introduced equation of state can be given a
string theory based motivation \cite{jackiw}. It has also been
generalized in different phenomenological ways \cite{berto}.
Another example for a unification scenario for the dark sector is
a bulk viscous model of the cosmic substratum \cite{winfried}.
While the Chaplygin gas model (in its traditional and generalized
forms) has been very successful in explaining the supernovae type
Ia data \cite{colistete}, there are claims that it does not pass
the tests connected with structure formation because of predicted
but not observed strong oscillations of the matter power
spectrum \cite{ioav}. It should be mentioned, however, that
oscillations in the Chaplygin gas component do not necessarily
imply corresponding oscillations in the observed baryonic power
spectrum \cite{avelino}.
\par
The generalized Chaplygin gas is characterized by the equation of state
\begin{equation}
\label{eos}
p = - \frac{A}{\rho^\alpha} \quad.
\end{equation}
For $A>0$ the pressure $p$ is negative, hence it may induce an accelerated
expansion of the universe. The corresponding sound speed is
positive as long as $\alpha
> 0$.
Recently, a gauge-invariant analysis of the baryonic matter power
spectrum for generalized Chaplygin gas cosmologies was shown to be
compatible with the data for parameter values $\alpha \approx 0$
and $\alpha \geq 3$ \cite{staro}. This result seems to strengthen
the role of Chaplygin gas type models as competitive candidates
for the dark sector. The present work provides a further
investigation along these lines. While we shall rediscover the
mentioned results of \cite{staro}, albeit in a different
framework, we also extend the scope of the analysis in the
following sense. The authors of \cite{staro} have shown that
Chaplygin gas cosmologies are consistent with the data from
structure formation for certain parameter configurations. Here we
ask additionally, whether or not the data really favor generalized
Chaplygin gases as unified models of the dark sector. By leaving
the density parameters of the Chaplygin gas and the
non-relativistic matter component, respectively, free, we allow
for a matter fraction that can be different from the pure baryonic
part. This is equivalent to hypothetically admit the existence of
an additional dark matter component. In other words, we do not
prescribe the unified model from the start. Moreover, our study is
not restricted to the spatially flat case.

Our study relies on a {\it neo-Newtonian} approach which
represents a major simplification of the problem. In some sense,
the neo-Newtonian equations can be seen as the introduction of a
first order relativistic correction to the usual Newtonian
equations \cite{harrison}. The neo-Newtonian equations for
cosmology \cite{mccrea,harrison,lima,reis} modify the  Newtonian
equations in a way that makes the pressure dynamically relevant
already for the homogeneous and isotropic background. This allows
us to describe an accelerated expansion of the Universe as the
consequence of a sufficiently large effective negative pressure in
a Newtonian framework. While the neo-Newtonian approach reproduces
the GR background dynamics exactly, differences occur at the
perturbative level. However, the GR first-order perturbation
dynamics and its neo-Newtonian counterpart coincide exactly in the
case of a vanishing sound speed \cite{reis}. On small scales one
expects the spatial pressure gradient term to be relevant and the
difference to the GR dynamics should be of minor importance. Since
the observational data correspond to modes that are well inside
the Hubble radius, the use of a Newtonian type approach seems
therefore adequate.

On this basis our analysis extends previous neo-Newtonian studies
to the two-component case. One of the components is a generalized
Chaplygin gas, the other one represents pressureless matter. The
advantage of employing a neo-Newtonian approach is a gain in
simplicity and transparency. Our neo-Newtonian approach reproduces
the parameter estimations for the unified dark matter/dark energy
in \cite{staro} also numerically. Backed up by this success of the
neo-Newtonian approach we then enlarge the scope of our analysis
and test the validity of the unified model itself by relaxing the
unified model priors used in \cite{staro}. Denoting the present
value of the Chaplygin gas density parameter by $\Omega_{c0}$, we
admit the total present matter density parameter $\Omega_{m0}$ to
be the sum of an additional dark matter component with density
parameter $\Omega_{dm0}$ and the baryon contribution
$\Omega_{b0}$, i.e., $\Omega_{m0} = \Omega_{dm0} + \Omega_{b0}$.
Leaving the density parameters free, we investigate whether or not
the unified model with $\Omega_{c0} \approx 0.96$, $\Omega_{b0}
\approx 0.04$ and $\Omega_{dm0} \approx 0$ is favored by the
large-scale structure data. We mention that a similar
investigation using supernova type Ia data reveals that the
unification scenario is the most favored one \cite{colistete}.
\par
In the framework of the neo-newtonian formalism, in the
conservation equation one takes into account the work done by the
pressure during the expansion of the universe. At the same time,
the equation for the gravitational potential must be modified in
order to render the equations compatible. This has been done in
references \cite{mccrea,harrison,lima}. The final equations are
\begin{eqnarray}
\frac{\partial\rho}{\partial t} + \nabla\cdot(\rho\vec v) + p\nabla\cdot\vec v &=& 0 \ ,\\
\frac{\partial\vec v}{\partial t} + \vec v\cdot\nabla\vec v &=& - \frac{\nabla p}{\rho + p} - \nabla\phi \ ,\\
\nabla^2\phi &=& 4\pi G(\rho + 3p) \ .
\end{eqnarray}
For the case of two non-interacting fluids with energy densities $\rho_c$ and $\rho_m$ and pressures
$p_c$ and $p_m = 0$, respectively, the equations are:
\begin{eqnarray}
\label{neo1}
\frac{\partial\rho_c}{\partial t} + \nabla\cdot(\rho_c\vec v_c) + p_c\nabla\cdot\vec v_c &=& 0 \ ,\\
\label{neo2}
\frac{\partial\vec v_c}{\partial t} + \vec v_c\cdot\nabla\vec v_c &=& - \frac{\nabla p_c}{\rho_c + p_c} - \nabla\phi \ ,\\
\label{neo3}
\frac{\partial\rho_m}{\partial t} + \nabla\cdot(\rho_m\vec u_m) &=& 0 \ ,\\
\label{neo4}
\frac{\partial\vec v_m}{\partial t} + \vec v_m\cdot\nabla\vec v_m &=& - \nabla\phi \ ,\\
\label{neo5} \nabla^2\phi &=& 4\pi G(\rho_m + \rho_c + 3p_c) \
.
\end{eqnarray}
The subscript $m$ stands for pressureless matter and the subscript
$c$ for the (generalized) Chaplygin gas component. Considering now
an isotropic and homogeneous universe with $\rho = \rho(t)$, $p =
p(t)$ and $\vec v = \frac{\dot a}{a}\vec r$, we find
\begin{eqnarray}
\biggr(\frac{\dot a}{a}\biggl)^2 + \frac{k}{a^2} &=& \frac{8\pi
G}{3}(\rho_m + \rho_c) \ , \\
\frac{\ddot a}{a} &=& - \frac{4\pi G}{3}(\rho_c + \rho_m + 3p_c)
\ .
\end{eqnarray}

Let us define the fractional density contrasts
\begin{equation}
\delta_c = \frac{\delta\rho_c}{\rho_c} \quad \mathrm{and }\quad \delta_m =
\frac{\delta\rho_m}{\rho_m} \
\end{equation}
for the Chaplygin gas and matter components, respectively, the
first-order perturbation equations for the system
(\ref{neo1})-(\ref{neo5}) are
\begin{eqnarray}
\ddot\delta_c &+& \biggr\{2\frac{\dot a}{a} -
\frac{\dot\omega_c}{1 + \omega_c} + 3\frac{\dot a}{a}(v_c^2 -
\omega_c)\biggl\}\dot\delta_c
+ \left\{3\biggr(\frac{\ddot a}{a} + \frac{\dot a^2}{a^2}\biggl)(v_c^2 - \omega_c) \qquad\qquad\right.\nonumber\\
&&\left. + 3\frac{\dot a}{a}\biggr[\dot v^2_c -
\dot\omega_c\frac{(1 + v_c^2)}{1 + \omega_c}\biggl] +
\frac{v_c^2\,k^2}{a^2} - 4\pi G\rho_c(1 + \omega_c)(1 +
3v_c^2)\right\}\delta_c = 4\pi G\rho_m(1 + \omega_c)\delta_m
\label{dddc}
\end{eqnarray}
and
\begin{equation}
\ddot\delta_m + 2\frac{\dot a}{a}\dot\delta_m - 4\pi
G\rho_m\delta_m = 4\pi G\rho_m(1 + 3v_c^2)\delta_c \ ,
\label{dddm}
\end{equation}
where $v_c^2 = \frac{\partial p_c}{\partial\rho_c}$ and $\omega_c
= \frac{p_c}{\rho_c}$. The quantity $k^{2}$ denotes the square of
the comoving wave vector. Dividing the equations (\ref{dddc}) and
(\ref{dddm}) by $H_0^2$ and redefining the time as
$t\,H_0\rightarrow t$, these equations become dimensionless. In
terms of the scale factor $a$ as dynamical variable, the system
(\ref{dddc})-(\ref{dddm}) takes the form
\begin{eqnarray}
\label{dddcbis} \delta''_c &+& \biggr\{\frac{2}{a} + g(a) -
\frac{\omega_c'(a)}{1 + \omega_c(a)} - 3\frac{1 +
\alpha}{a}\omega_c(a)\biggl\}\delta'_c
\nonumber\\
&&\quad - \biggr\{3\biggr[\frac{g(a)}{a} + \frac{1}{a^2}\biggl](1
+ \alpha)\omega_c(a) + \frac{3}{a}\biggr(\frac{1 + \alpha}{1 +
\omega_c(a)}\biggl)\omega'_c(a) +
\frac{\alpha\omega_c(a)\,k^2l_H^2}{a^2\,f(a)}
\nonumber\\
&&\quad + \frac{3}{2}\frac{\Omega_{c0}}{f(a)}h(a)[1 + \omega_c(a)][1 - 3\alpha\omega_c(a)]\biggl\}\delta_c
= \frac{3}{2}\frac{\Omega_{m0}}{a^3\,f(a)}[1 + \omega_c(a)]\delta_m \
\end{eqnarray}
and
\begin{equation}
\delta''_m + \biggr[\frac{2}{a} + g(a)\biggl]\delta'_m -
\frac{3}{2}\frac{\Omega_{m0}}{a^3\,f(a)}\delta_m =
\frac{3}{2}\frac{\Omega_{c0}}{f(a)}h(a)[1 -
3\alpha\omega_c(a)]\delta_c \ ,\label{dddma}
\end{equation}
where the quantity $k^{2}$ denotes the square of the comoving wave
vector and $l_H = c\,H_{0}^{-1 }$ is the present Hubble radius.
The prime denotes a derivative with respect to $a$ and the
definitions
\begin{eqnarray}
f(a) &=& \frac{\dot a^2}{H_{0}^{2}} = \left[\frac{\Omega_{m0} + \Omega_{c0}a^3\,h(a)}{a} + \Omega_{k0}\right] \ ,\\
g(a) &=& \frac{\ddot a}{\dot a^2} = - \frac{\Omega_{m0} + \Omega_{c0}[h(a) - 3\bar A\, h^{-\alpha}]a^3}{2a[\Omega_{m0} + \Omega_{c0}a^3h(a) + \Omega_{k0}a]} \ ,\\
h(a) &=& [\bar A + (1 - \bar A)a^{-3(1 + \alpha)}]^\frac{1}{1 + \alpha} \ , \\
\omega_c(a) &=& - \frac{\bar A}{h(a)^{1 + \alpha}} \  ,
\end{eqnarray}
with
\begin{equation}
\bar A = \frac{A}{\rho_{c0}^{1 + \alpha}} \ , \quad v_{s0}^2 = \alpha\bar A \
\end{equation}
have been used. Recall that $\Omega_{m0} = \Omega_{dm0} +
\Omega_{b0}$. For the unified model to be an adequate description
one expects $\Omega_{m0} \approx \Omega_{b0}$. In case the data
indicate a substantial fraction of $\Omega_{dm0}$, the unified
model will be disfavored.

The power spectrum is defined by
\begin{equation}
{\cal P} = \delta_k^2 \quad ,
\end{equation}
where $\delta_k$ is the Fourier transform of dimensionless density
contrast $\delta_m$. We will constrain the free parameters using the quantity
\begin{equation}
\chi^2 = \sum_i\biggr(\frac{{\cal P}_i^o - {\cal P}_i^t}{\sigma_i}\biggl)^2
\quad ,
\label{chi}
\end{equation}
where ${\cal P}_i^o$ is the observational value for the power
spectrum, ${\cal P}_i^t$ is the corresponding theoretical result and
$\sigma_i$ denotes the error bar. The index $i$ refers to a measurement
corresponding to given wavenumber. The quantity (\ref{chi}) qualifies
the fitting of the observational data for a given theoretical
model with specific values of the free parameters.
Hence, $\chi^2$ is a function of the free parameters
of the model. The probability distribution function is then
defined as
\begin{equation}
F(x_n) = F_0\,e^{-\chi^2(x_n)/2} \quad ,
\end{equation}
where the $x_n$ denote the ensemble of free parameters and $F_0$
is a normalization constant. In order to obtain an estimation for
a given parameter one has to integrate (marginalize) over all the
other ones. For a more detailed description of this statistical
analysis see reference \cite{colistete}. From now on we focus on
the 2dFGRS observational data for the power spectrum \cite{cole}.
We use the data that are related with the linear approximation,
that is, those for which $k\,h^{-1} \leq 0.185 Mpc^{-1}$, where
$h$ is defined by $H_{0} \equiv 100\cdot h \mathrm{km/s\cdot
Mpc}$. This definition should not be confused with the preceding
definition of the function $h(a)$. To fix the initial conditions
we use the BBKS transfer function \cite{bbks}. This procedure is
described in more detail in references \cite{sola,saulo}.
\par
\begin{center}
\begin{figure}[!t]
\begin{minipage}[t]{0.225\linewidth}
\includegraphics[width=\linewidth]{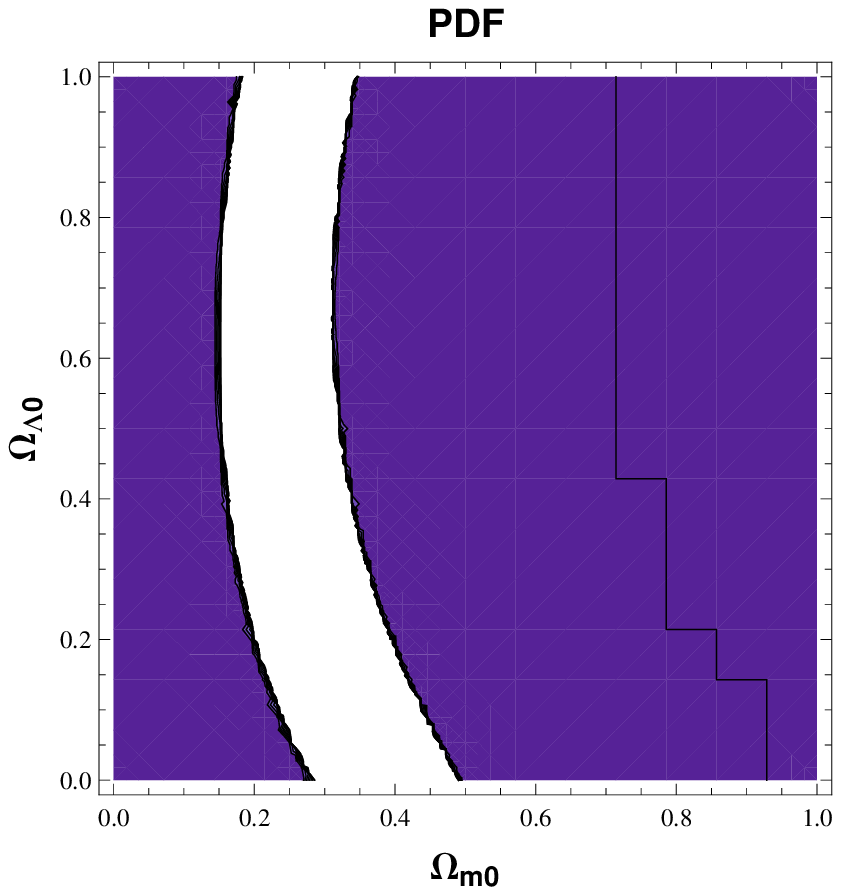}
\end{minipage} \hfill
\begin{minipage}[t]{0.225\linewidth}
\includegraphics[width=\linewidth]{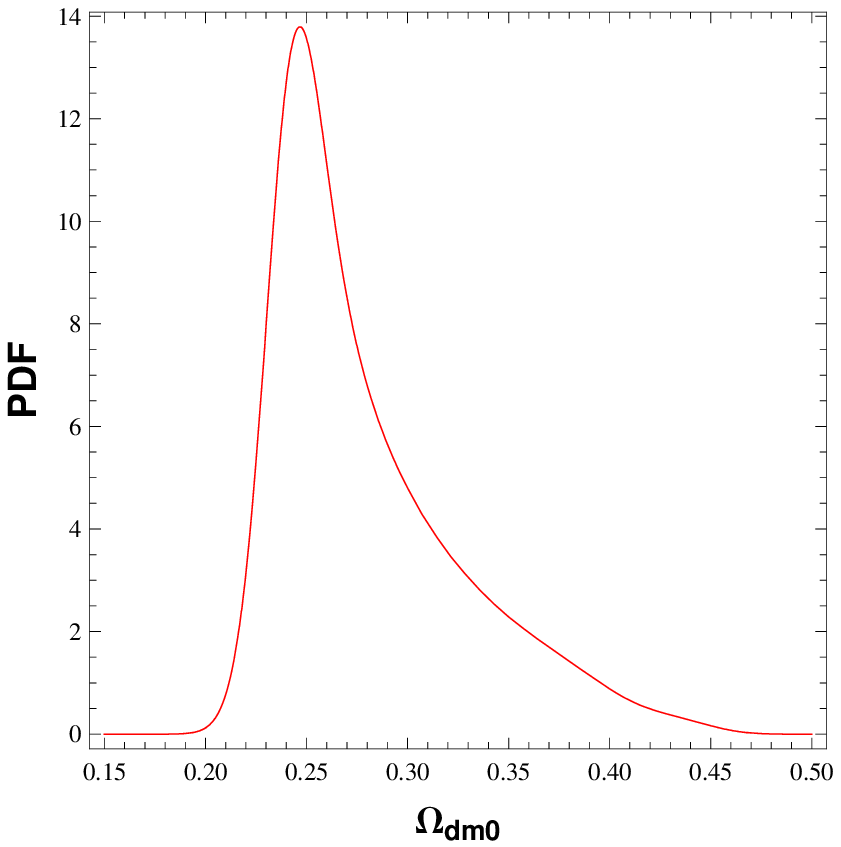}
\end{minipage} \hfill
\begin{minipage}[t]{0.225\linewidth}
\includegraphics[width=\linewidth]{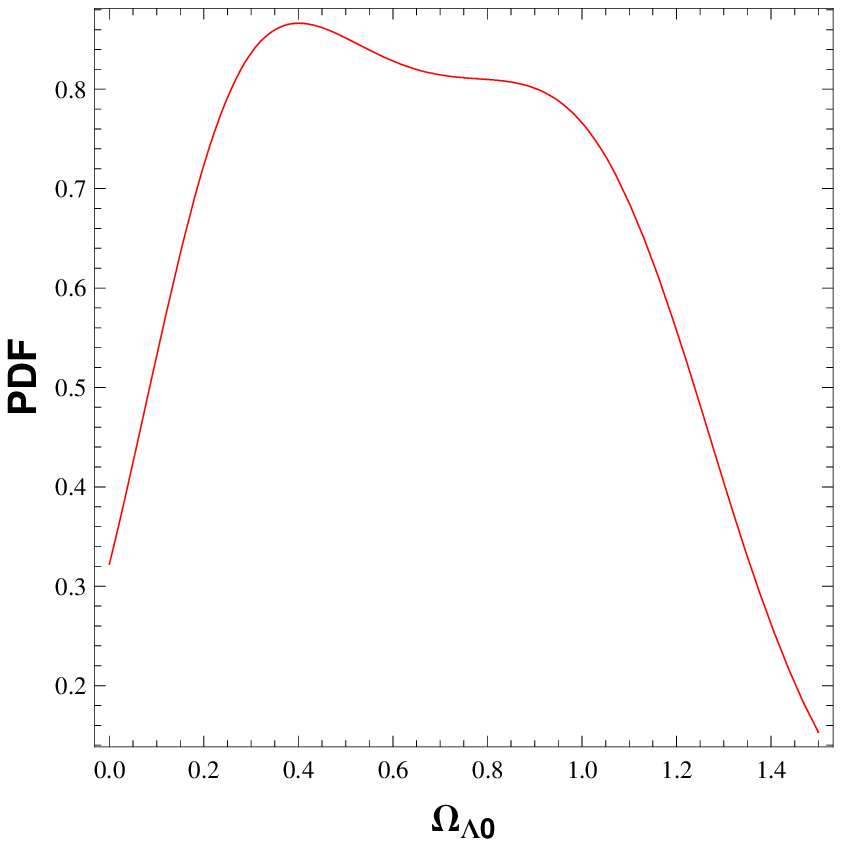}
\end{minipage} \hfill
\caption{{\protect\footnotesize The two-dimensional probability
distribution function (PDF) for $\Omega_{dm0}$ and
$\Omega_{\Lambda0}$ (left) and the corresponding one-dimensional
probability distribution functions for the non-flat $\Lambda$CDM
model. In the left panel: the darker  the color, the smaller the
probability.}} \label{LCDM}
\end{figure}
\end{center}

To ``gauge" our approach, let us first consider the $\Lambda$CDM
model. In the general (non-flat) case there are two parameters:
$\Omega_{dm0}$ and $\Omega_{\Lambda0}$. In figure \ref{LCDM}   we
show the two-dimensional probability distribution function (PDF)
as well the one-dimensional PDFs for the dark matter parameter
$\Omega_{dm0}$ and for the cosmological constant parameter
$\Omega_{\Lambda0}$, respectively. From the two dimensional
graphic it is clear that there is a large degeneracy for the
parameter $\Omega_{\Lambda0}$, while the region of allowable
values for $\Omega_{dm0}$ is quite narrow. The degeneracy for the
cosmological constant density is less visible in the
one-dimensional PDF graphic, but it is still considerable.
Incidentally, the minimum value for the $\chi^2$ parameter is
$0.3822$ for $\Omega_{dm0} = 0.2387$ and $\Omega_{\Lambda0} =
0.5937$, corresponding to an open universe.
\par
\begin{center}
\begin{figure}[!t]
\begin{minipage}[t]{0.225\linewidth}
\includegraphics[width=\linewidth]{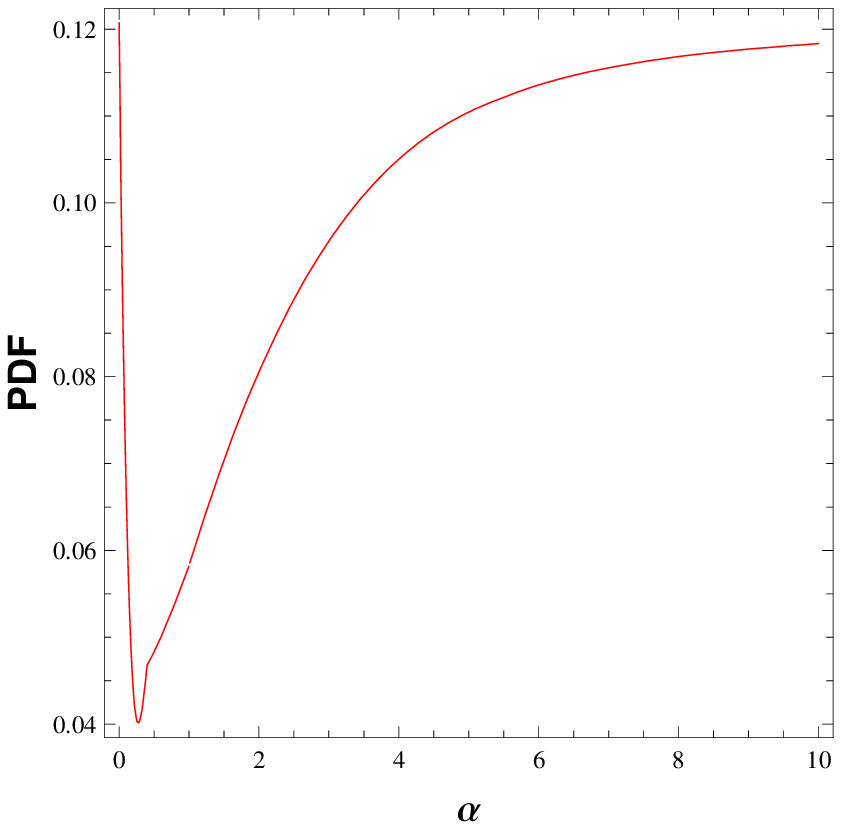}
\end{minipage} \hfill
\begin{minipage}[t]{0.225\linewidth}
\includegraphics[width=\linewidth]{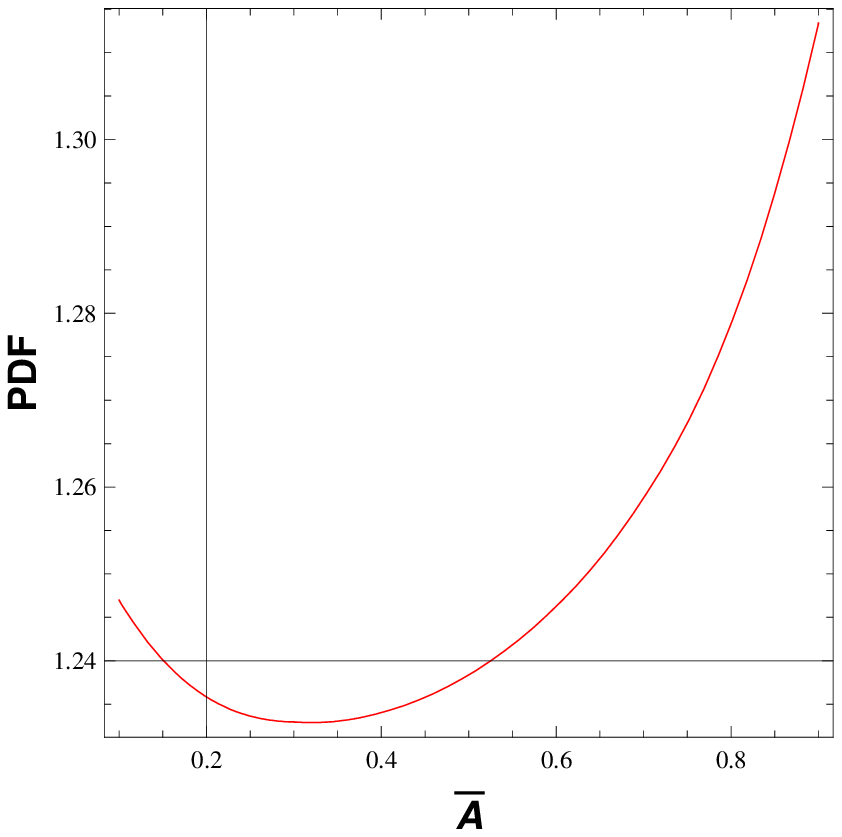}
\end{minipage} \hfill
\begin{minipage}[t]{0.225\linewidth}
\includegraphics[width=\linewidth]{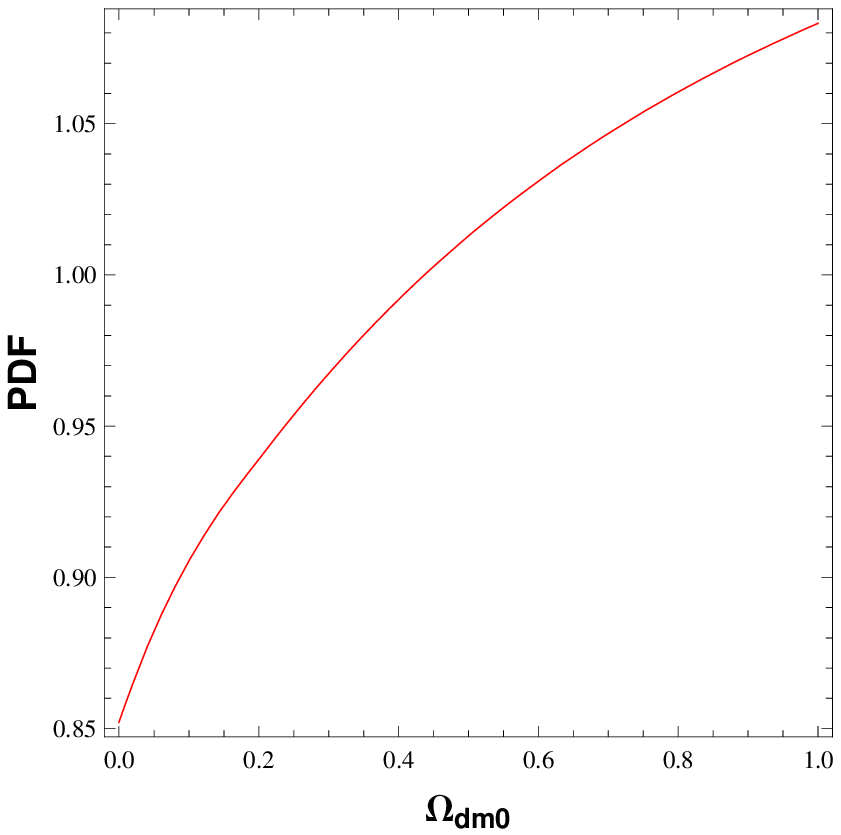}
\end{minipage} \hfill
\begin{minipage}[t]{0.225\linewidth}
\includegraphics[width=\linewidth]{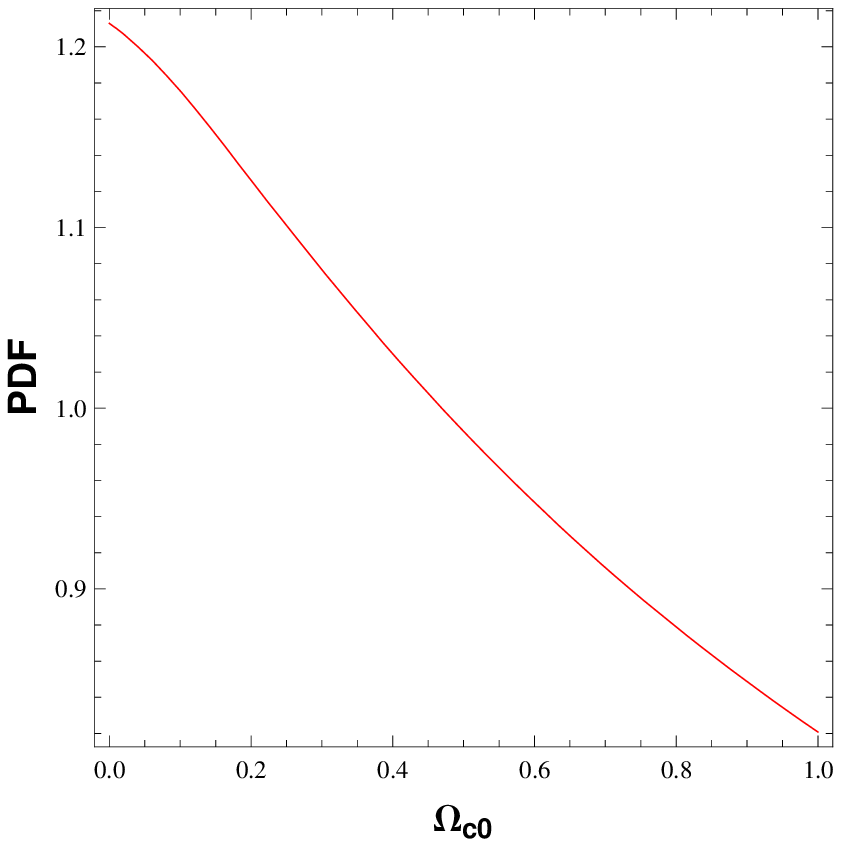}
\end{minipage} \hfill
\caption{{\protect\footnotesize The results for the general case
with four free parameters. From left to right: the one-dimensional
PDFs  for $\alpha$, $\bar A$, $\Omega_{c0}$ and $\Omega_{dm0}$.}}
\label{4p}
\end{figure}
\end{center}

The four free parameters to be constrained in our Chaplygin gas
model are $\Omega_{dm0}$, $\Omega_{c0}$, $\bar A$ and $\alpha$.
The one-dimensional PDFs for $\alpha$, $\bar A$, $\Omega_{dm0}$
and $\Omega_{c0}$ are displayed in figure \ref{4p}. It can be seen
that the preferred values are either $\alpha \ll 1$ or $\alpha
\geq 2$, while the probability is higher for large values of
$\Omega_{dm0}$ and small values of $\Omega_{c0}$. This show
clearly that the unification scenario is disfavored.
\begin{center}
\begin{figure}[!t]
\begin{minipage}[t]{0.225\linewidth}
\includegraphics[width=\linewidth]{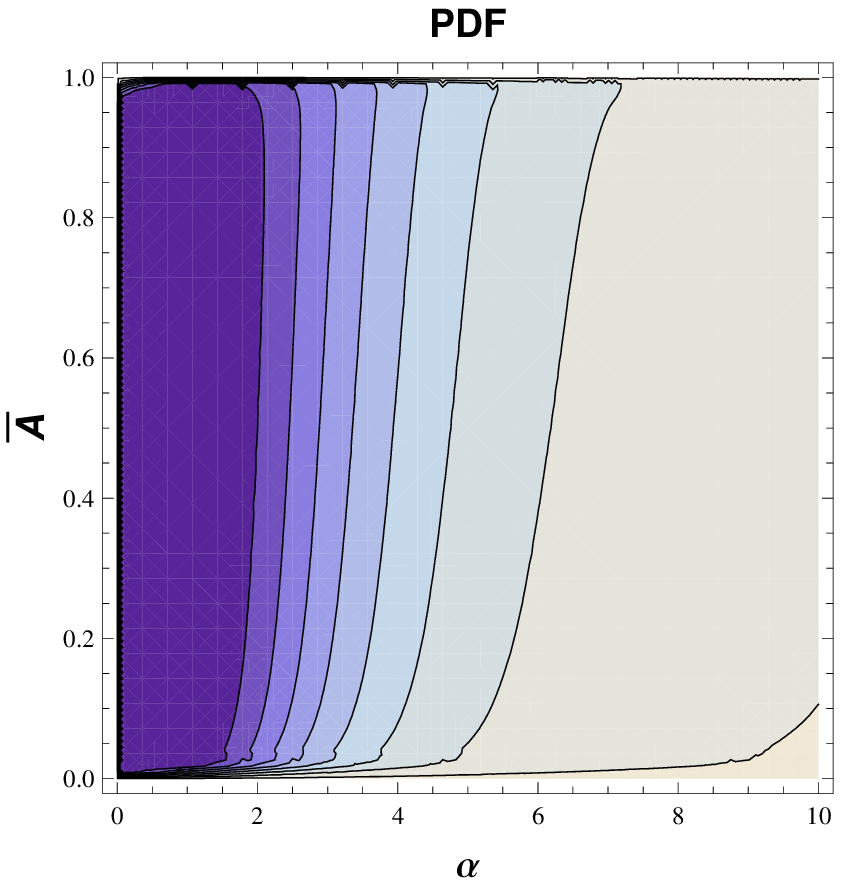}
\end{minipage} \hfill
\begin{minipage}[t]{0.225\linewidth}
\includegraphics[width=\linewidth]{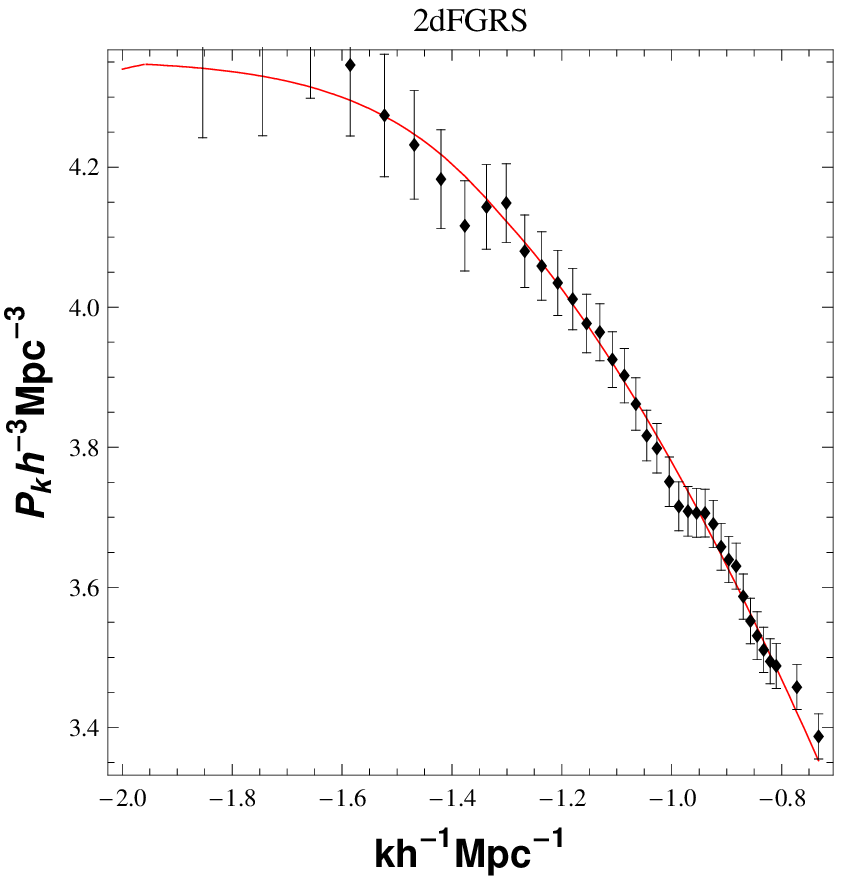}
\end{minipage} \hfill
\begin{minipage}[t]{0.225\linewidth}
\includegraphics[width=\linewidth]{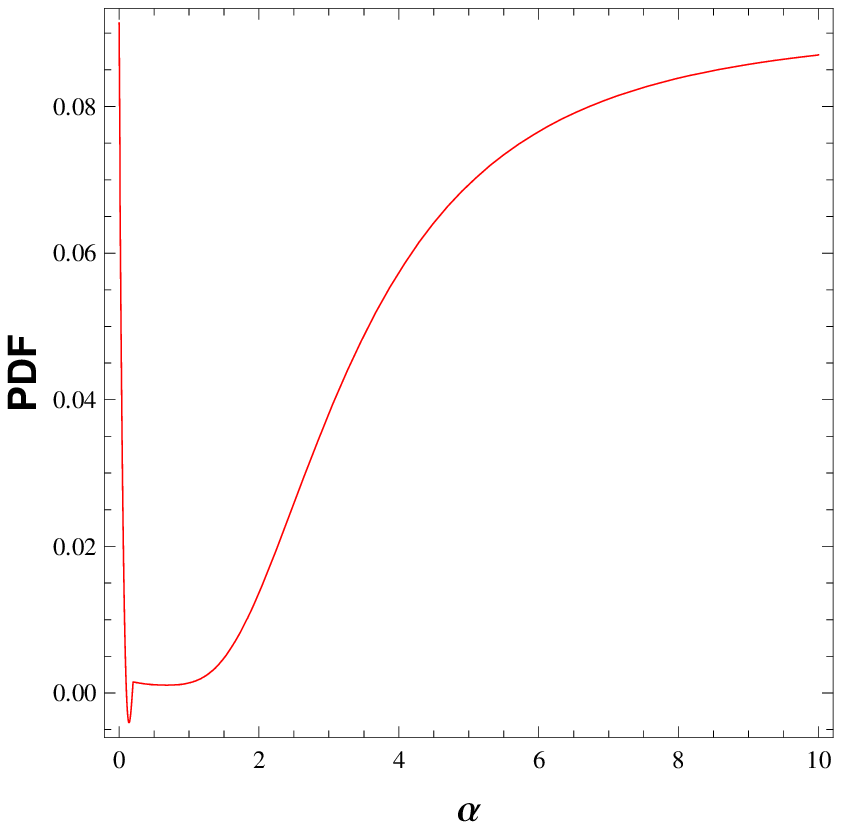}
\end{minipage} \hfill
\begin{minipage}[t]{0.225\linewidth}
\includegraphics[width=\linewidth]{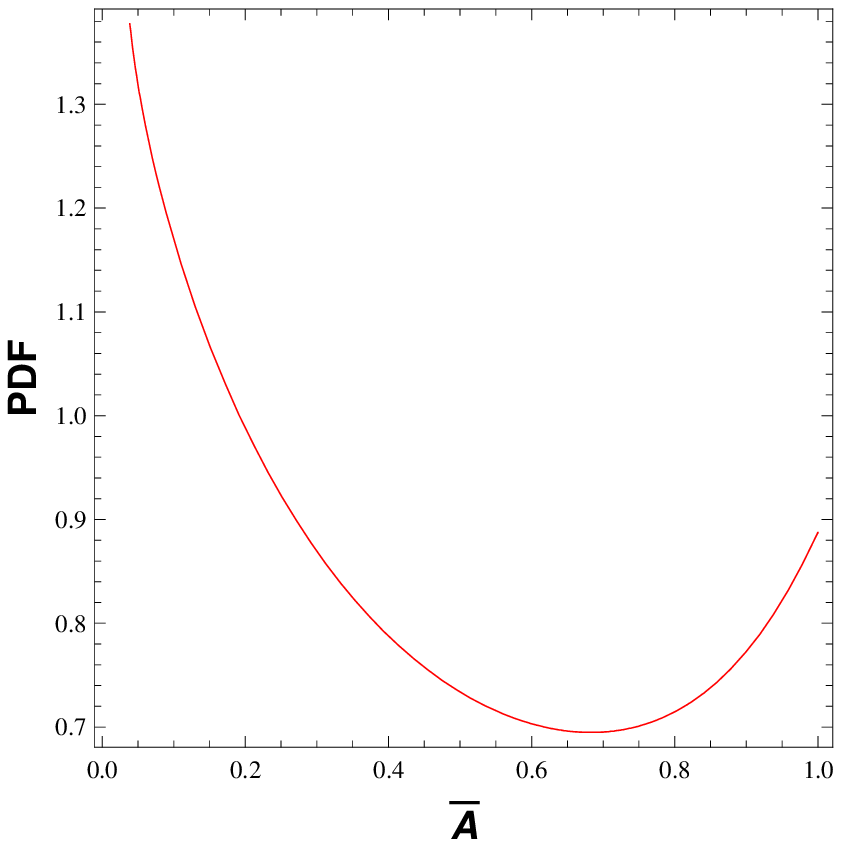}
\end{minipage} \hfill
\caption{{\protect\footnotesize The results for the flat case with
$\Omega_{b0} = 0.043$, $\Omega_{dm0} = 0$ and $\Omega_{c0} =
0.957$, corresponding to the unification scenario. From
left to right: the two-dimensional PDF for $\alpha$ and $\bar A$
(with the same color convention as before), the best fitting curve
for the power spectrum, the one-dimensional PDFs for $\alpha$ and
$\bar A$. }} \label{UniPrior}
\end{figure}
\end{center}
\par
If the unification scenario with dark matter and dark energy as a
single fluid in a spatially flat universe is imposed from the
beginning, the results of reference \cite{staro} are essentially
confirmed: there are parameter ranges for which the data are well
described by the generalized Chaplygin gas model, see figure \ref{UniPrior}. The probability
distribution function for $\alpha$ is high for very small (near
zero) or very large (greater than 2) values of $\alpha$. Allowing
the parameter $\bar A$ to vary, we find that its one-dimensional
PDF initially decreases with $\bar A$, but increases as $\bar A =
1$ is approached. Notice that values $\alpha > 1$ imply a
superluminal sound speed and are therefore unphysical (see,
however, \cite{staro}).
\par
What is the origin of these apparently contradictory results? The
first aspect to be mentioned is that the matter power spectrum
data only poorly constrain the dark energy component. Even for the
$\Lambda$CDM model the matter power spectrum gives information
mainly on the dark matter component, the dark energy component
remaining largely imprecise. It is not by chance that the dark
energy concept emerged from the supernova data. Our results for
the Chaplygin gas model show  that a large amount of dark matter,
different from those described by the Chaplygin gas, is necessary
to fit the data. However, the dispersion is quite high. For the
flat case with a three-dimensional parameter space we find at
$2\sigma$, that $\Omega_{dm0} = 1^{+0.00}_{-0.91}$. Another point
is the use of the neo-Newtonian formalism. However, for small
values of the parameter $\alpha$, the main case of interest
here,
the differences to the full general relativistic treatment
are not expected to be substantial. Moreover, in the cases of
overlap the results of the full theory are reproduced.  Finally,
possible statistical subtleties may influence the outcome of the
investigation. But as far as we could test the statistical
analysis (precision, crossing different information, etc), the
results seem to be robust. If this is really the case, we must
perhaps live with the fact that, while the SNe type Ia data favor
a unified model of the dark sector \cite{colistete}, this scenario
is disfavored if large scale structure data are taken into
account, unless specific priors are imposed.

\vspace{1.0cm}

{\bf Acknowledgement}.  We thank FAPES and CNPq (Brazil) for
partial financial support (grants 093/2007 (CNPq and FAPES) and EDITAL
FAPES No. 001/2007). J.C.F. thanks also the french-brazilian
scientific cooperation CAPES/COFECUB and the Institut of
Astrophysique de Paris (IAP) for the kind hospitality during the
elaboration of this work.


\begin{thebibliography}{99}

\bibitem{moschella} A.Y. Kamenshchik, U. Moschella and V. Pasquier, Phys. Lett.
{\bf B511}, 265(2001).

\bibitem{jackiw} R. Jackiw, {\it A particle field theorist's lectures on supersymmetric, non abelian fluid mechanics and d-branes},
physics/0010042.

\bibitem{berto}
M.C. Bento, O. Bertolami and A.A. Sen, Phys. Rev. {\bf D66},
043507 (2002).

\bibitem{winfried} R. Colistete Jr., J.C. Fabris, J. Tossa and W.
Zimdahl, Phys. Rev. {\bf D76}, 103516(2007).

\bibitem{colistete} R. Colistete Jr, J. C. Fabris, S.V.B. Gon\c{c}alves and P.E. de Souza, Int. J. Mod. Phys. {\bf D13}, 669(2004);
R. Colistete Jr., J. C. Fabris and S.V.B. Gon\c{c}alves, Int. J.
Mod. Phys. {\bf D14}, 775(2005); R. Colistete Jr. and J. C.
Fabris, Class. Quant. Grav. {\bf 22}, 2813(2005).

\bibitem{ioav} H. Sandvik, M. Tegmark, M. Zaldarriaga and I. Waga,
Phys. Rev. {\bf D69}, 123524(2004).

\bibitem{avelino} L.M.G. Be\c{c}a, P.P. Avelino, J.P.M. de
Carvalho and C.J.A.P. Martins, Phys. Rev. {\bf D67}, 101301
(2003).

\bibitem{staro} V. Gorini, A.Y. Kamenshchik, U. Moschella, O.F.
Piatella and A.A. Starobinsky, JCAP {\bf 02}, 016 (2008).

\bibitem{bbks} J.M. Bardeen, J.R. Bond, N. Kaiser and A.S. Szalay,
Astrophys. J. {\bf 304}, 15 (1986); J. Martin, A. Riazuelo and M.
Sakellariadou, Phys. Rev. {\bf D61}, 083518 (2000).

\bibitem{sola} J.C. Fabris, I.L. Shapiro and J. Sol\`a, JCAP {\bf
0712}, 007 (2007).

\bibitem{saulo} H.A. Borges, S. Carneiro, J.C. Fabris and C.
Pigozzo, Phys. Rev. {\bf D77}, 043513 (2008).

\bibitem{mccrea} W.H. McCrea, Proc. R. Soc. London {\bf 206}, 562
(1951).

\bibitem{harrison} E.R. Harrison, Ann. Phys. (N.Y.) {\bf 35}, 437
(1965).

\bibitem{lima} J.A.S. Lima, V. Zanchin and R. Brandenberger,
Month. Not. R. Astron. Soc. {\bf 291}, L1(1997).

\bibitem{reis} R.R.R. Reis, Phys. Rev. {\bf D67}, 087301 (2003);
erratum-ibid {\bf D68}, 089901(2003).

\bibitem{cole} S. Cole et al., Month. Not. R. Astron. Soc. {\bf
362}, 505 (2005).


\end{thebibliography}
\end{document}